\documentclass[final]{iaus}
\usepackage{graphicx}
\usepackage{natbib}

\title[JD 11.~~Aug. 24, 2002 CME] 
{The August 24, 2002 Coronal Mass Ejection: \\
When a Western Limb Event Connects to Earth}

\author[Lugaz, Roussev \& Sokolov]   
{No{\'e} Lugaz$^1$, Ilia I. Roussev$^1$ \and Igor V. Sokolov$^2$}

\affiliation{$^1$Institute for Astronomy, University of Hawaii, 2680 Woodlawn Dr.,
Honolulu, HI, 96822, USA \\ email: {\tt nlugaz@ifa.hawaii.edu, iroussev@ifa.hawaii.edu} \\[\affilskip]
$^2$ Department of AOSS, University of Michigan, 2455 Hayward St., Ann Arbor, MI 48198 \\email: {\tt igorosk@umich.edu}}

\pubyear{2008}
\volume{IAU257}  
\pagerange{1-7}
\jname{Universal Heliophysical Processes}
\editors{N. Gopalswamy, D. Webb \& A. Nindos, eds.}
\begin{document}

\maketitle

\begin{abstract}
We discuss how some coronal mass ejections (CMEs) originating from the western limb of the Sun are associated with space weather effects such as solar energetic particles (SEPs), 
shock or geo-effective ejecta at Earth. We focus on the August 24, 2002 coronal mass ejection, a fast ($\sim$ 2000 km~s$^{-1}$) eruption originating from W81. Using a three-dimensional 
magneto-hydrodynamic simulation of this ejection with the Space Weather Modeling Framework (SWMF), we show how a realistic initiation mechanism enables us to study the deflection of the CME in the corona and the heliosphere. Reconnection of the erupting magnetic field with that of neighboring streamers and active regions modify the solar connectivity of the field lines connecting to Earth and can also partly explain the deflection of the eruption during the first tens of minutes. 
Comparing the results at 1 AU of our simulation with observations by the {\it ACE} spacecraft, we find that the simulated shock does not reach Earth, but has a maximum angular span of about 120$^\circ$, and reaches 35$^\circ$ West of Earth in 58~hours. We find no significant deflection of the CME and its associated shock wave in the heliosphere, and we discuss the consequences for the shock angular span. 
\end{abstract}

\firstsection

\section{Introduction}
\subsection{The Eruption on August 24, 2002}
On August 24, 2002, active region (AR) 10069 was near the western limb of the Sun (W81) when it produced a powerful (X3.1) flare associated with a fast and wide coronal mass ejection (CME). 
This event has been well studied due to extensive observations remotely by SoHO/LASCO and SoHO/UVCS \citep[]{Raymond:2003} as well in-situ by the {\it Wind} and {\it ACE} spacecraft among others and also 
its inclusion as one of the Solar Heliospheric Interplanetary Environment (SHINE) campaign events. 
Based on LASCO observations, this was an instance of wide and fast CME, with an average speed of 1,900~km~s$^{-1}$ within the first 20 solar radii. 
This event has been mostly studied in association with another wide and fast CME on April 21, 2002 for its association with a large Solar Energetic Particle (SEP) event \citep[]{Tylka:2005, Tylka:2006}. Based on an increased
of the iron-to-oxygen ration at large energy, these authors proposed that the shock geometry explain this difference: quasi-perpendicular for the August 24 and quasi-parallel for the April 21 one. The presence or absence of a 
reflecting boundary at or slightly ahead of Earth associated with a previous eruption has also been recently proposed to explain these difference \citep[]{Tan:2008}.

\subsection{Studying Western-Limb Ejections}
Western limb events such as the August 24 and April 21, 2002 CMEs present a number of challenges for space weather prediction. Due to the Parker's spiral, the Earth is on average magnetically connected with regions at the solar surface around W55 (for a 400~km~s$^{-1}$ background wind). Therefore, SEP events are preferentially associated with western events. Events such as the April 21, 2002 and August 24, 2002 present additional challenges since the SEP arrival time at Earth corresponds to a particle release height of less than 5~$R_\odot$. If one believes that these particles are accelerated by the CME-driven shock wave, there are a number of scenarios to explain the observations. First, the shock must have formed low in the corona, and then, within 
5~$R_{\odot}$ of the solar surface, it either spans at least 60$^\circ$, or is smaller but it must have been significantly deflected towards the east, or the magnetic field line connecting Earth to the solar surface significantly diverges from the nominal Parker's spiral. Such differences of up to 30$^\circ$ during SEP events between flare cites and the magnetic footpoint of the Earth on the solar surface have been reported before \citep[]{Ippolito:2005}.

Western limb events are often associated with shock and sometimes ejecta at 1~AU. Both August 24 and April 21, 2002 were associated with a shock wave at 1~AU which transited in about 55~hrs. Among large geomagnetic storms (Dst $\le$ -100~nT) from the past solar cycle \citep[]{Zhang:2008}, at least 6 were caused by a shock at Earth associated with an ejection western of W73. This fact, again, seems to imply either a very large span of the shock wave, a large deflection of the CME, or a combination of both. 

Until now, it has been hard to study observationally the deflection of a CME in the corona or in the heliosphere due to the paucity of observations, especially in the near-Earth environment. \citet{Tripathi:2004} reported deflection up to 20$^\circ$ within the first hour after the initiation of an eruption based on a series of LASCO images. Shock span can also be estimated with white-light images, but it is limited by assumptions of symmetry and geometrical effects \citep[]{Cremades:2006}. In these two examples, the determination of the CME span and deflection can be only be made for limb CMEs and only in the meridional direction. Additionally, the longitudinal extent of shocks can be estimated  from multiple-spacecraft measurements; but until the launch of STEREO, it could only be done with spacecraft at different heliospheric distances, for example by the Helios spacecraft in the 1980s \citep[]{DeLucas:2008}. The launches of STEREO and SMEI have also made 3-D tomography of CME easier \citep[]{Jackson:2002}. On the theoretical and numerical sides, previous studies have focused on the deflection of a CME in the heliosphere due to its interaction with the solar wind \citep[]{Wang:2004}. 
Here, based on a 3-D numerical simulation of the August 24, 2002 event, we discuss how, by using a new and realistic model of solar eruptions associated with a realistic model of the coronal magnetic field, we can study these different effects. 
 
\begin{figure}[t]
\begin{minipage}[]{1.0\linewidth}
\begin{center}
{\includegraphics*[width=6cm]{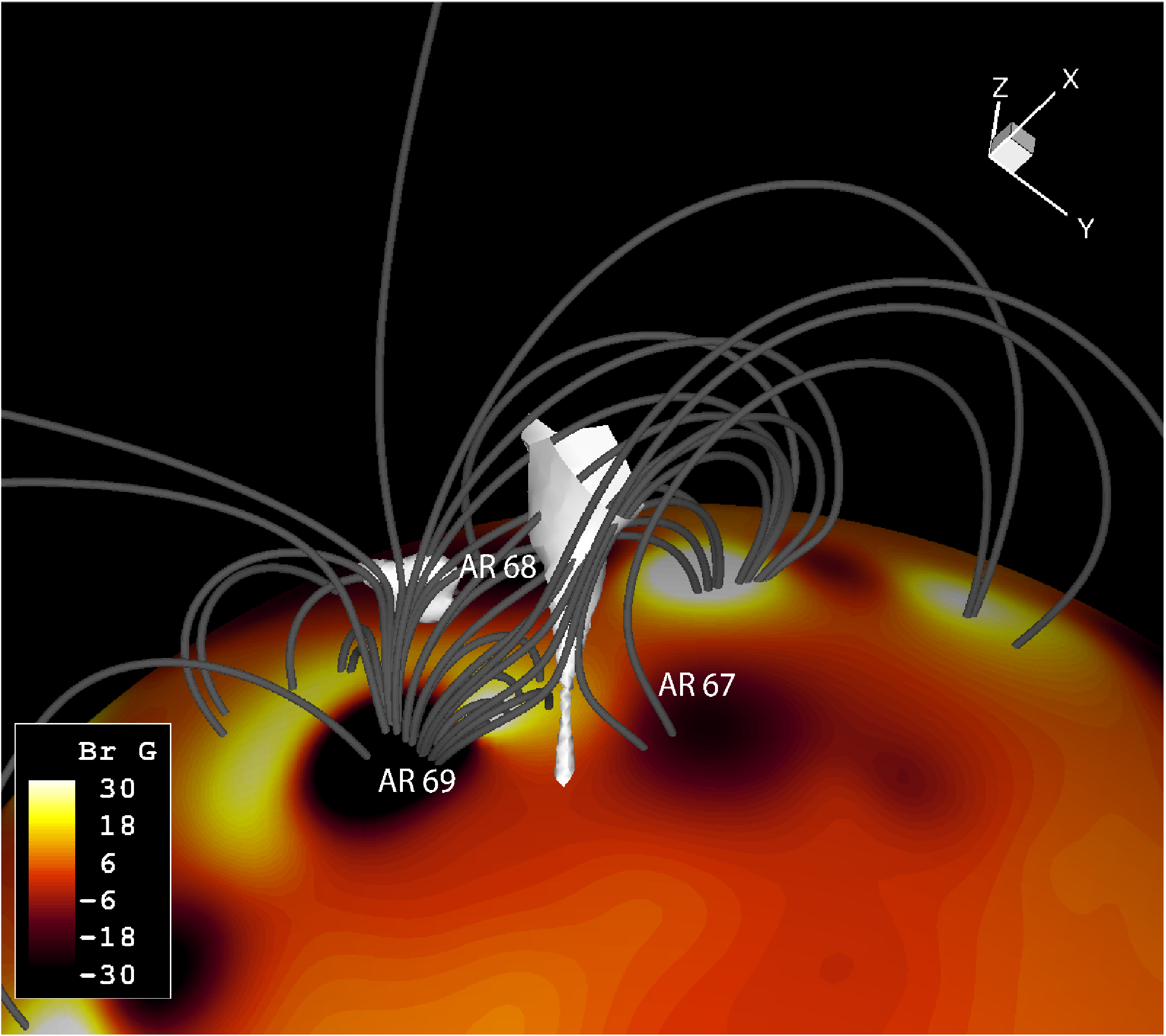}} 
{\includegraphics*[width=6cm]{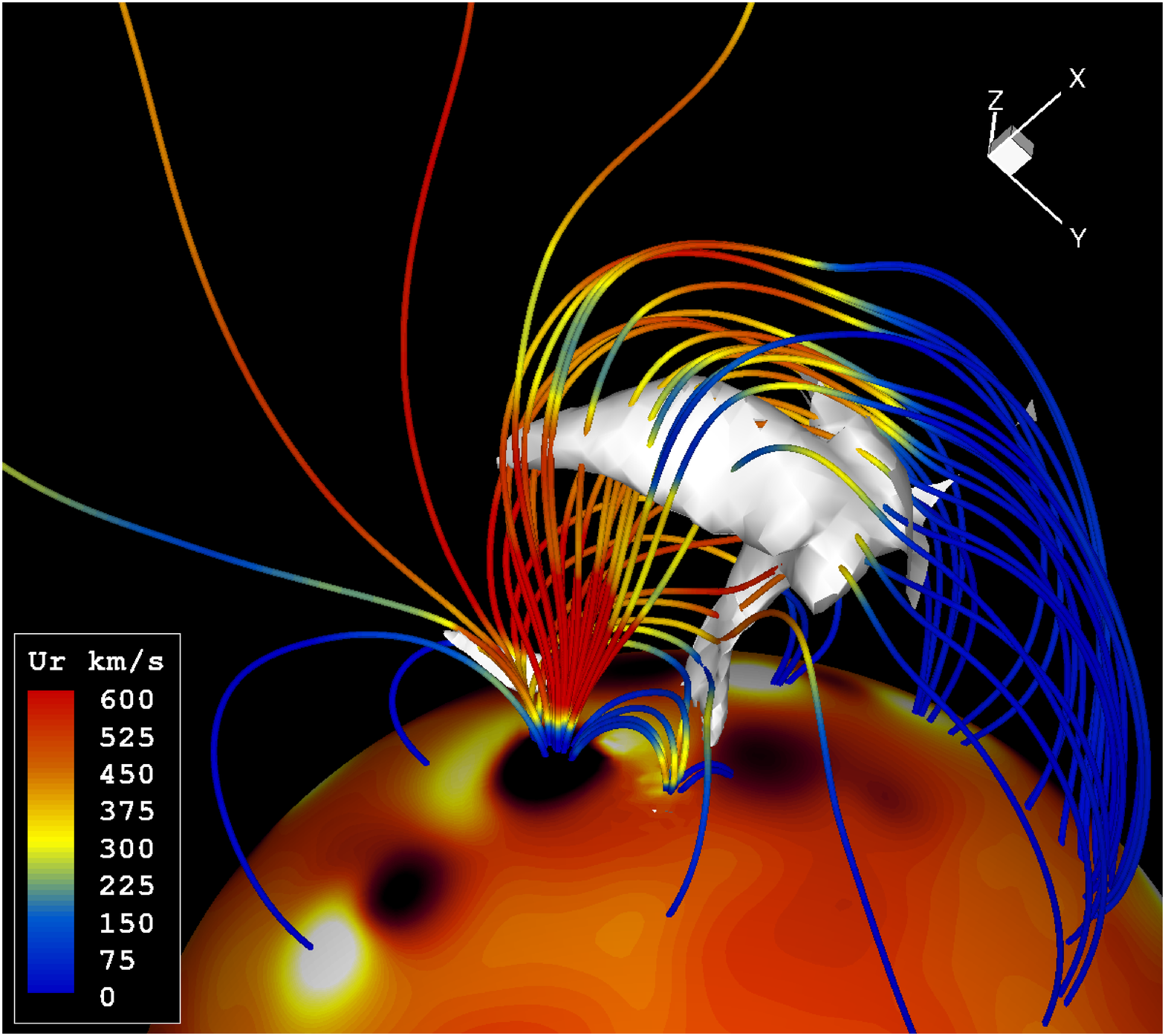}}
\end{center}
\end{minipage}\hfill
\caption{Magnetic topology before (\textit{left}) and after (\textit{right}, $t = 20$min) the shearing phase. The solar surface is color-coded with the radial magnetic field strength and the 
magnetic field lines with the radial velocity for the right panel. The null point and quasi-separatrix layer (QSL)
are visualized as white isosurfaces of plasma beta equal to 0.15 \textit{left}) and 0.4 (\textit{right}). }
\label{fig:Topology}
\end{figure}
 
\section{Simulation Setup}
\subsection{Numerical Domains}

In our numerical model, the steady-state solar corona and solar
wind are constructed following the methodology of \cite{Roussev:2003}, further
described in \cite{Roussev:2007}. The initial
condition for the coronal magnetic field is calculated by means of potential
field extrapolation, following \cite{Altschuler:1977}, with boundary condition for
the radial magnetic field at the Sun, $B_R$, provided by full-disk SoHO/MDI
observations taken four days before the eruption when
the AR was closer to the disk center and better observed.
The ``solar'' boundary in our model is placed at a height of 0.1~$R_{\odot}$
above the photosphere.  
The plasma parameters are prescribed
in an ad-hoc manner, through a variable polytropic index, in order to mimic the
physical properties of streamers and coronal holes once a steady-state (non-potential) is reached. 

The time-dependent MHD equations for a single compressible fluid are solved using
the Space Weather Modeling Framework \citep[]{Toth:2005} using two physical domain:
the Solar Corona (SC):, $\{-20
\le x \le 20, -20 \le y \le 20, -20 \le z \le 20 \} \, R_{\odot}$ and Inner Heliosphere (IH): $\{-220
\le x \le 220, -220 \le y \le 220, -220 \le z \le 220 \} \, R_{\odot}$.
We prescribe the initial grid in a way such as AR 69 is resolved with cells as small as $4.9 \times
10^{-3} \, R_{\odot}$,  4 times finer than the rest of the solar surface. Additionally, the radial direction 
above the active region (resp. direction of the field lines connecting to Earth) is refined with cells of 
0.08~$R_{\odot}$ up to 8 (resp. 5)~$R_{\odot}$ and 0.16~$R_{\odot}$ up to 14 (resp. 10)~$R_{\odot}$.
The total number of computational cells is of the order of 1.77 and 14.66 millions, with largest meshes of
size $1.25~R_{\odot}$ and $3.44~R_{\odot}$ for SC and IH, respectively.

To the initial magnetic field constrained by SoHO/MDI data, we superimposed newly emerged magnetic flux
simulated by a dipolar magnetic field of two point charges. These two charges are initially separated by $5 \times 10^3$~km 
and buried at a depth of $3 \times 10^4$~km under the solar surface, and chosen so that the peak value of the radial magnetic field 
at the solar surface is about 47~Gauss.

 \begin{figure}[t]
\begin{minipage}[]{1.0\linewidth}
\begin{center}
{\includegraphics*[width=6cm]{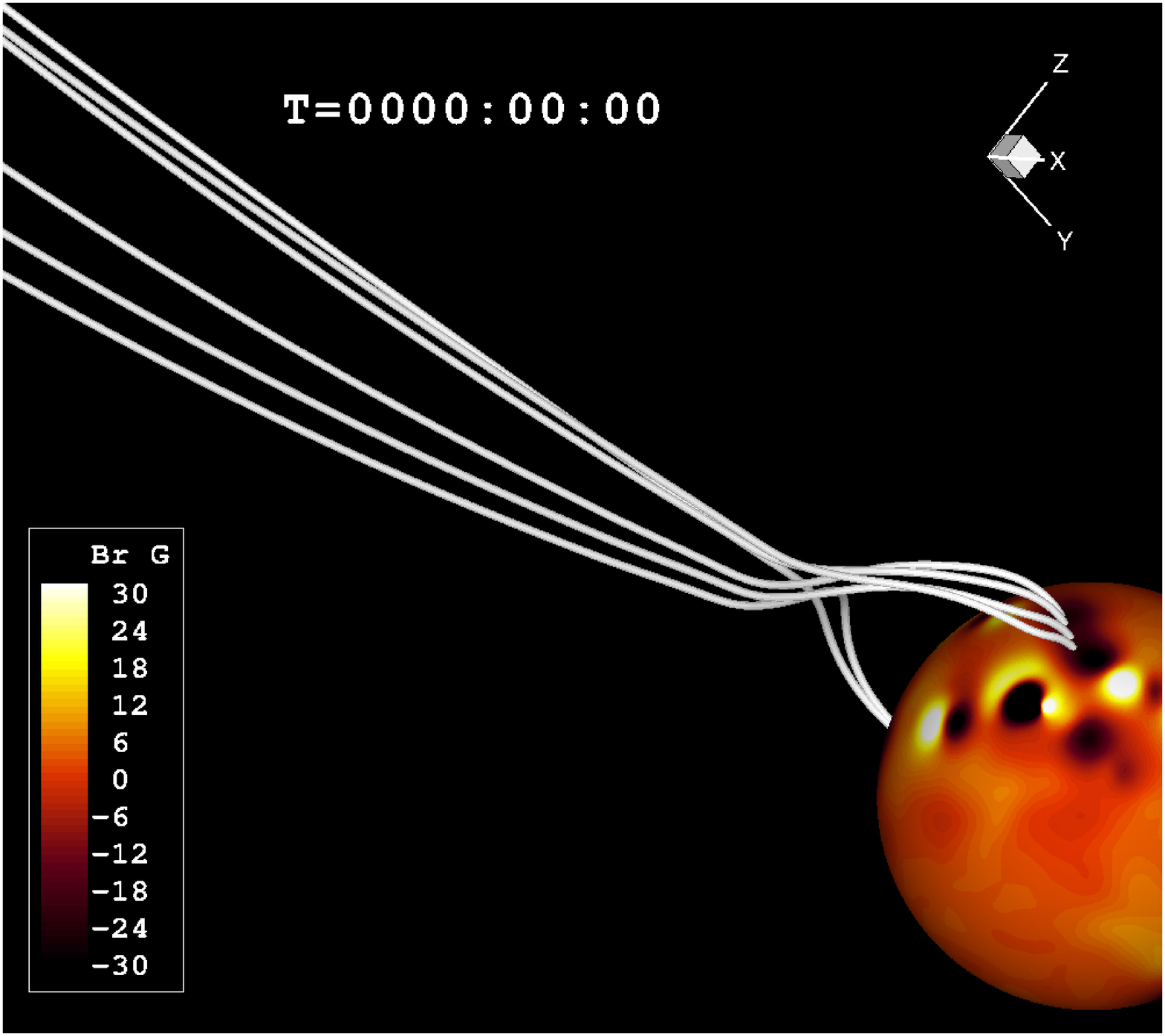}} 
{\includegraphics*[width=6cm]{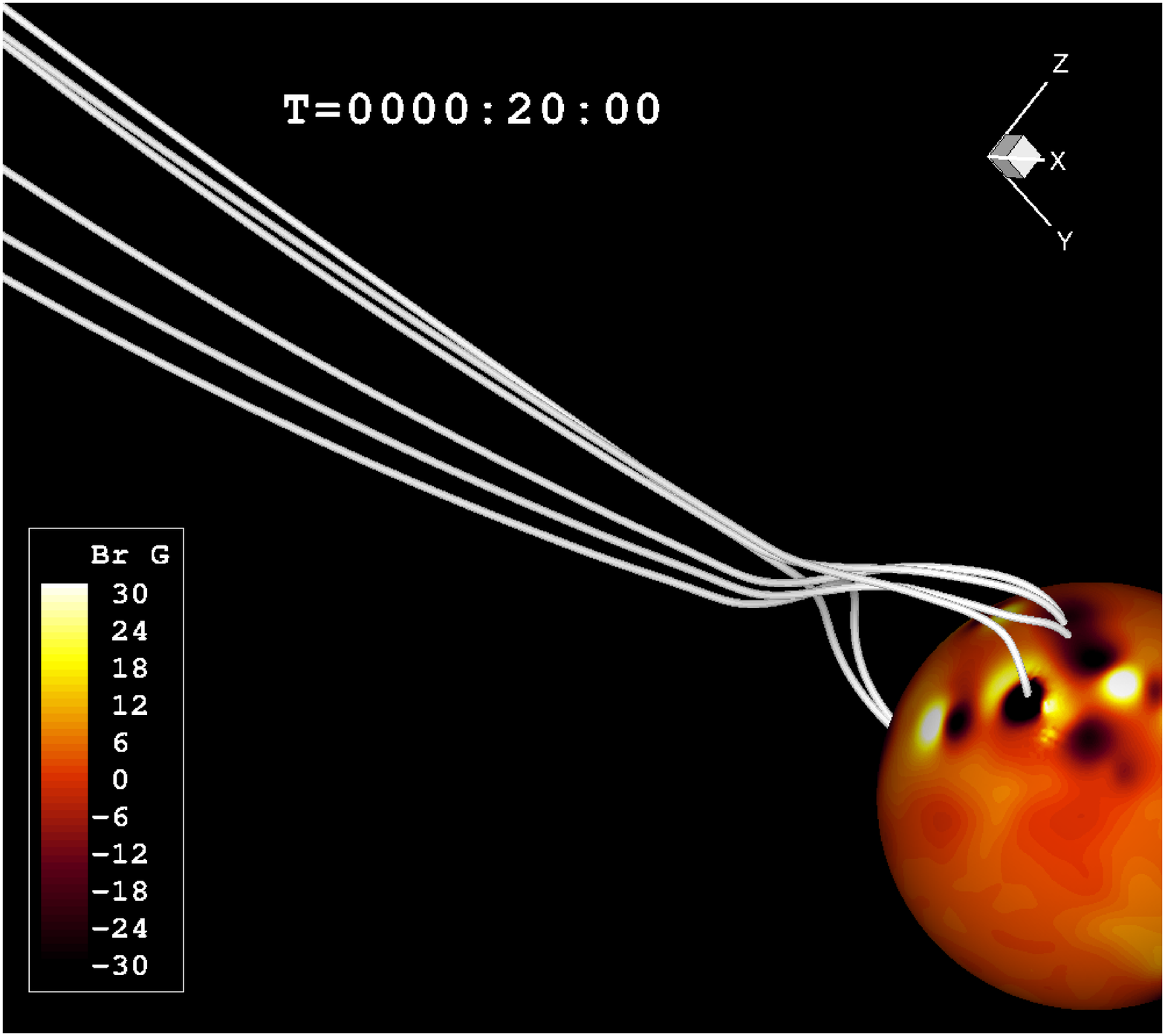}}\\
{\includegraphics*[width=6cm]{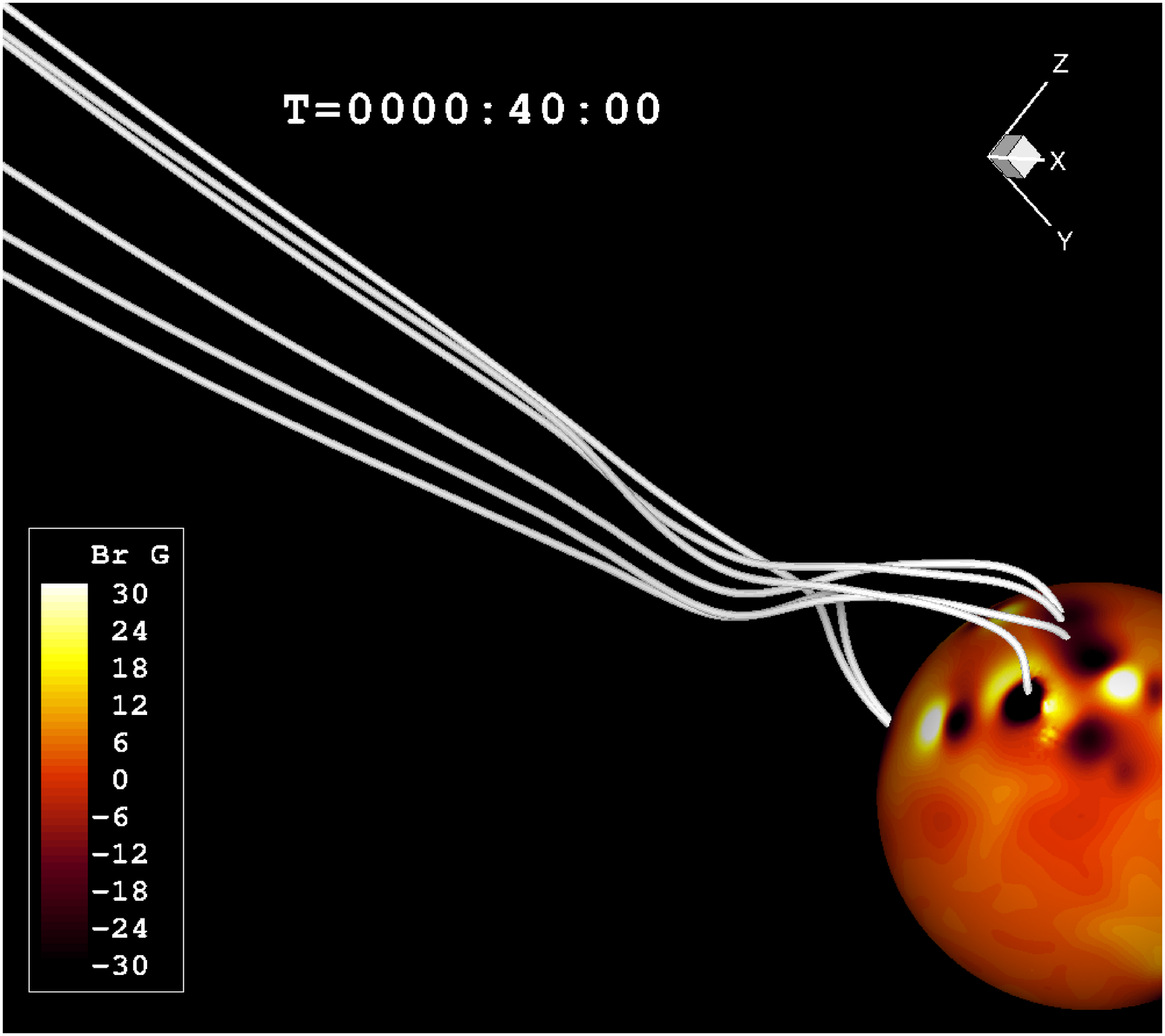}} 
{\includegraphics*[width=6cm]{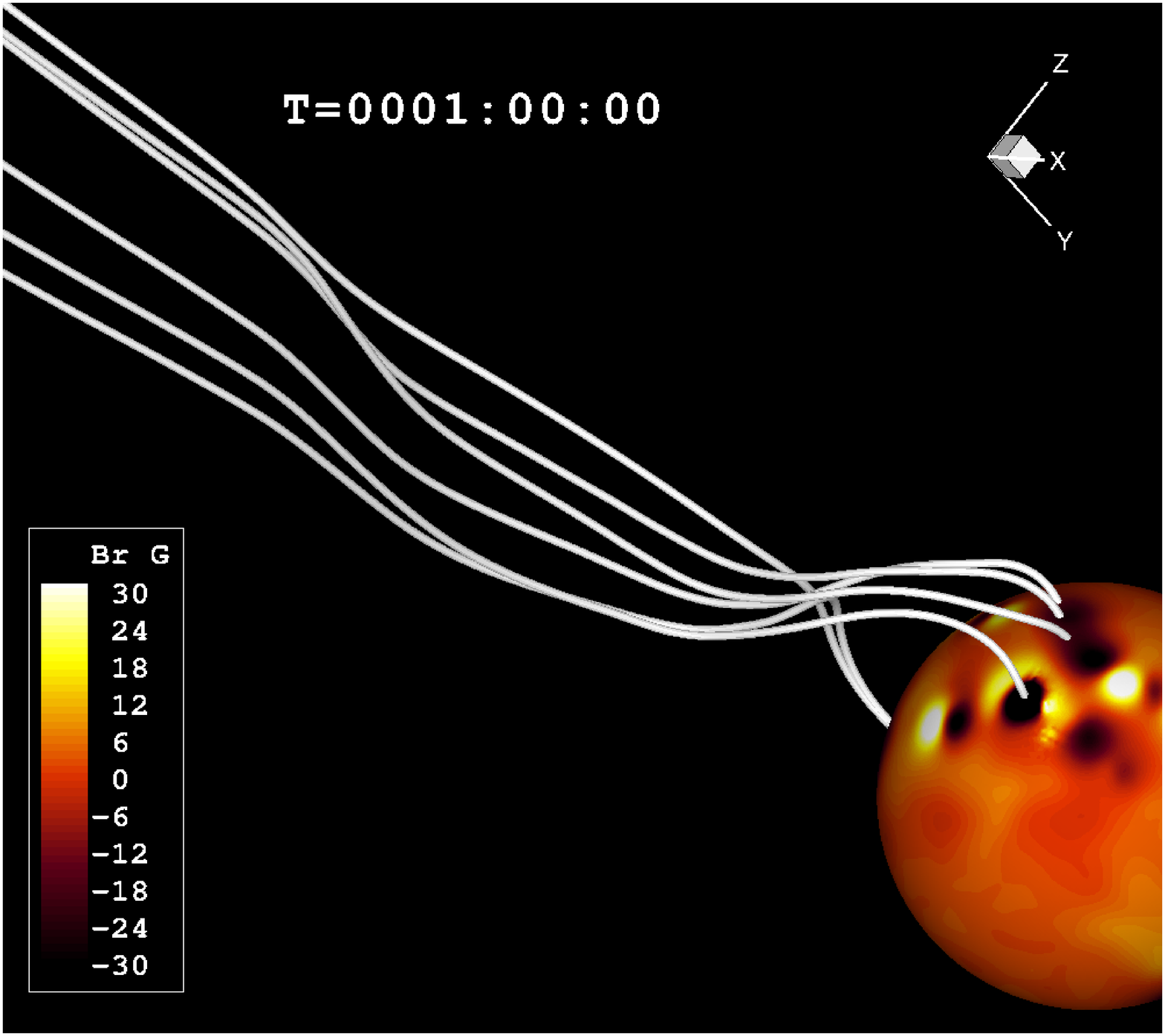}}
\end{center}
\end{minipage}\hfill
\caption{Field lines connecting to the vicinity of Earth and their evolution during the early phase of the eruption. The solar surface is color-coded with the radial magnetic field strength. 
Note the change of connectivity of one of the field lines at $t = 20$~min which connect AR 10069 to Earth. Note also the propagation of the shock wave along the field lines at later times.}
\label{fig:connection}
\end{figure}

\subsection{Solar Eruption Model}

To initiate the eruption, we use the model described in \cite{Roussev:2007}.
To summarize, once the steady-state is reached at $t=0$, the two magnetic charges are
moved apart quasi-steadily up to $t=t_S=20$~min with a speed which is ramped up in $t_S/3$ to
80~km~s$^{-1}$; the charge motion is stopped at $t=t_S$.  

In addition to updating the radial component of the dipole field at the boundary, 
we also impose the accompanying horizontal boundary motions.  
As the result of moving the charges apart, the magnetic field lines connecting the two spots
of the dipole are stretched.  
With appropriate choice of parameters
describing the relative position of the charges, their strength and speed of
motion, one can achieve a quasi-steady magnetic field evolution toward a state
that is no longer stable. Then, as the result of loss of confinement with the
overlying field, the ``energized'' magnetic field of the dipole erupts, manifesting a CME.  

\section{Eruption and Coronal Evolution}
\subsection{Loss of Equilibrium}
One of the main results of the work in \citet{Roussev:2007} was to recognize the importance of the pre-existing magnetic topology in the initiation of the eruption. This is first and foremost 
because reconnection at the pre-existing null points and quasi-separatrix layers (QSLs) enables the sheared and energized magnetic flux of the dipole to erupt. 
This work was the continuation and adaptation to realistic background coronal magnetic fields of previous works by \citet{Antiochos:1999, Galsgaard:2005} and \citet{Pontin:2007}.
There are two main pre-existing topological features important to understand this eruption: a null point between ARs 10067 and 10069 and a QSL
between ARs 10066, 10068 and 10069 (the magnetic topology before the shearing phase is illustrated in the left panel of Figure~1). Noteworthy is the fact that there are open field lines originating from AR 10069 and AR 10067 which 
pass at proximity of the null point. 

During the shearing, current builds up along the loops
connecting the two magnetic spots of the dipole and the magnetic field lines expand until they reconnect with overlaying field through the QSL. The QSL is disrupted and becomes a current sheet
which starts erupting (as illustrated in the right panel of Figure~1). The erupting field lines are now connecting AR 10069 to ARs 10067 and 10066. As some of these field lines expand further, 
they reconnect though the null point and some of them open up. The main motion of the erupting flux is radially above the initial position of the QSL but the reconnection through the null point between ARs 69 and 67 also enables 
the expansion of the CME towards solar east (Earth direction).

\begin{figure}[t]
\begin{minipage}[]{1.0\linewidth}
\begin{center}
{\includegraphics*[width=6cm]{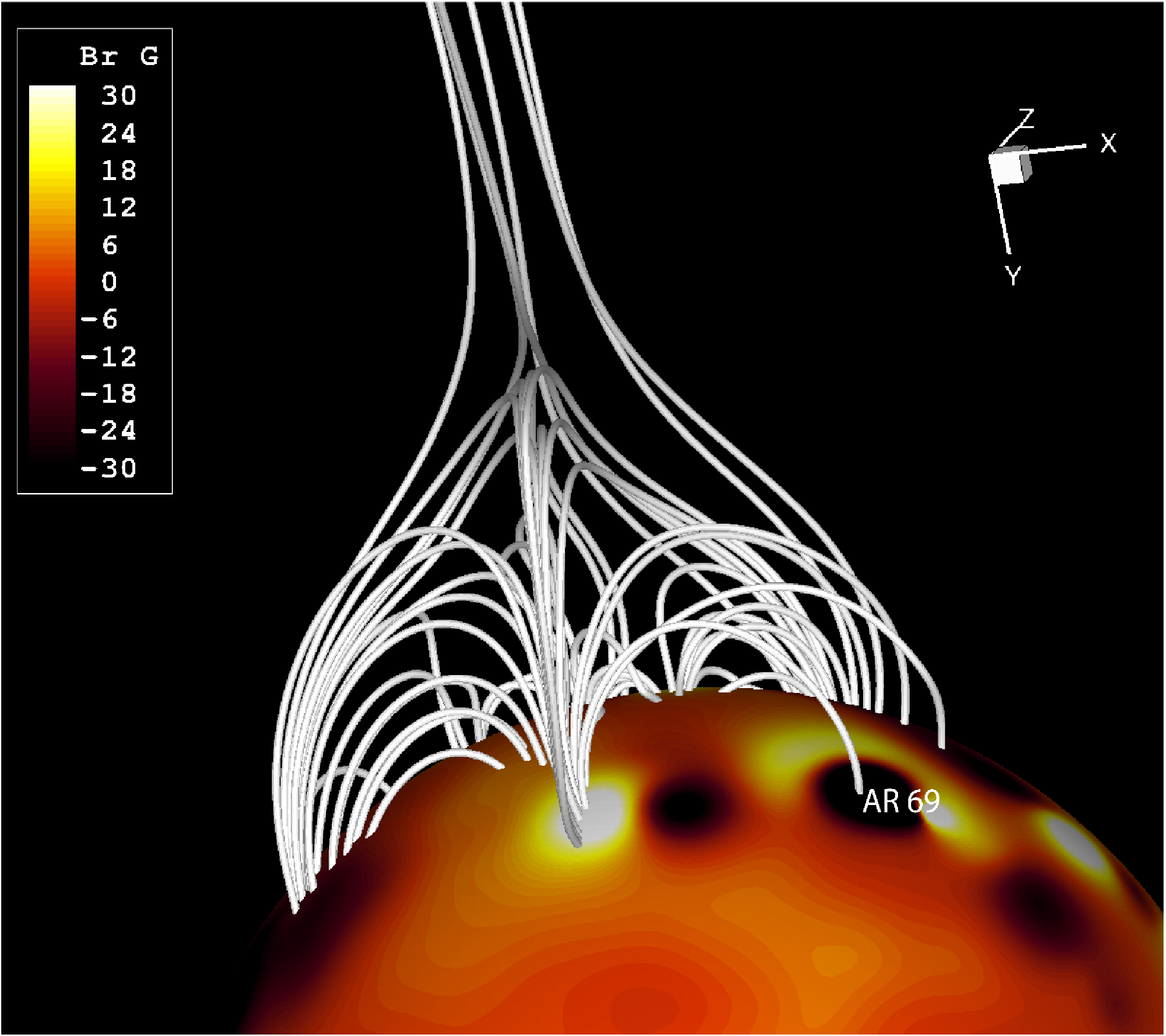}} 
{\includegraphics*[width=6cm]{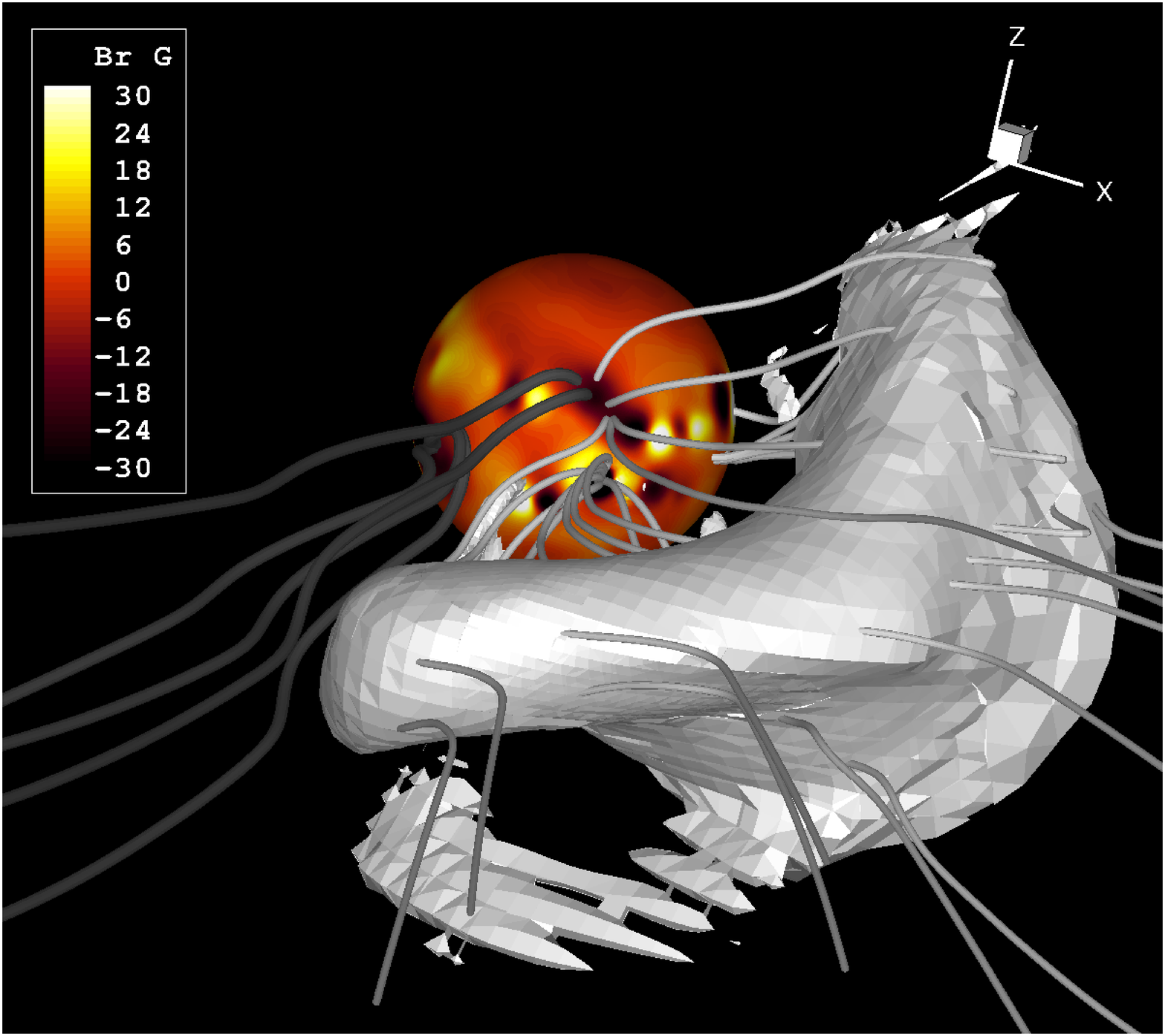}}
\end{center}
\end{minipage}\hfill
\caption{{\it Left}: Solar footpoints of the magnetic field lines connecting to the vicinity of Earth, which span from S15W05 to N20W75. {\it Right}: Visualization of the shock wave 1 hour after the start of the shearing phase. Field lines connecting to Earth's vicinity are shown with darker grey and larger radii. The white surface is an isosurface of Aflv{\'e}nic Mach equal to 1.}
\label{fig:Shock}
\end{figure}

\subsection{Magnetic Connection to Earth}

The August 24, 2002 eruption was associated with a large Solar Energetic Particle event, even though the magnetic connectivity to Earth is hard to assess due to the position of the AR 81$^\circ$ west of disk center. The timing of the arrival
of SEPs at Earth is consistent with particle release at less than 5~$R_\odot$. Additionally, contrary to our model which is not potential, the potential field source surface model \citep[]{Altschuler:1977} shows that there are no open magnetic field lines originating from AR~69. Our simulation can help explaining why a SEP event was indeed observed at Earth. We will not focus on a given field line, but on a set of field lines spanning about $10^\circ$ at 1~AU including the one connected to Earth on August 24, 2002 01UT. A number of reasons make the determination of the exact footpoint of the field line connected to Earth difficult and compelled us to consider a stack of field lines instead. First, the solar magnetic field is reconstructed from observations on August 21, 2002, 3 days prior to the studied event and the coronal and photospheric magnetic fields have certainly changed in this time span. Second, as noted by \citet{Tan:2008}, the presence of prior ejections in the heliosphere may significantly modify the magnetic connectivity. There was a number of ejections prior to the August 24 ones, alhough there was no clear ejecta passing Earth at this time according to satellite data. Random walk of the field lines may also account for up to 10$^\circ$ longitudinal variation on the solar surface \citep[]{Ippolito:2005}. Last, as can be seen on the top left panel of Figure~2, field lines connecting to the vicinity of Earth before the start of the shearing phase have footprints in two very distinct zones on the solar surface through a QSL: one around N20W70 and one around S15W10. The ``average'' position of the field lines connecting to Earth is about W40, indeed not departing too much from the nominal Parker's spiral. However, the magnetic topology close to the Sun is such that some of these field lines connect to the proximity of AR~69.

\begin{figure}[t]
\begin{minipage}[]{1.0\linewidth}
\begin{center}
{\includegraphics*[width=6.6cm]{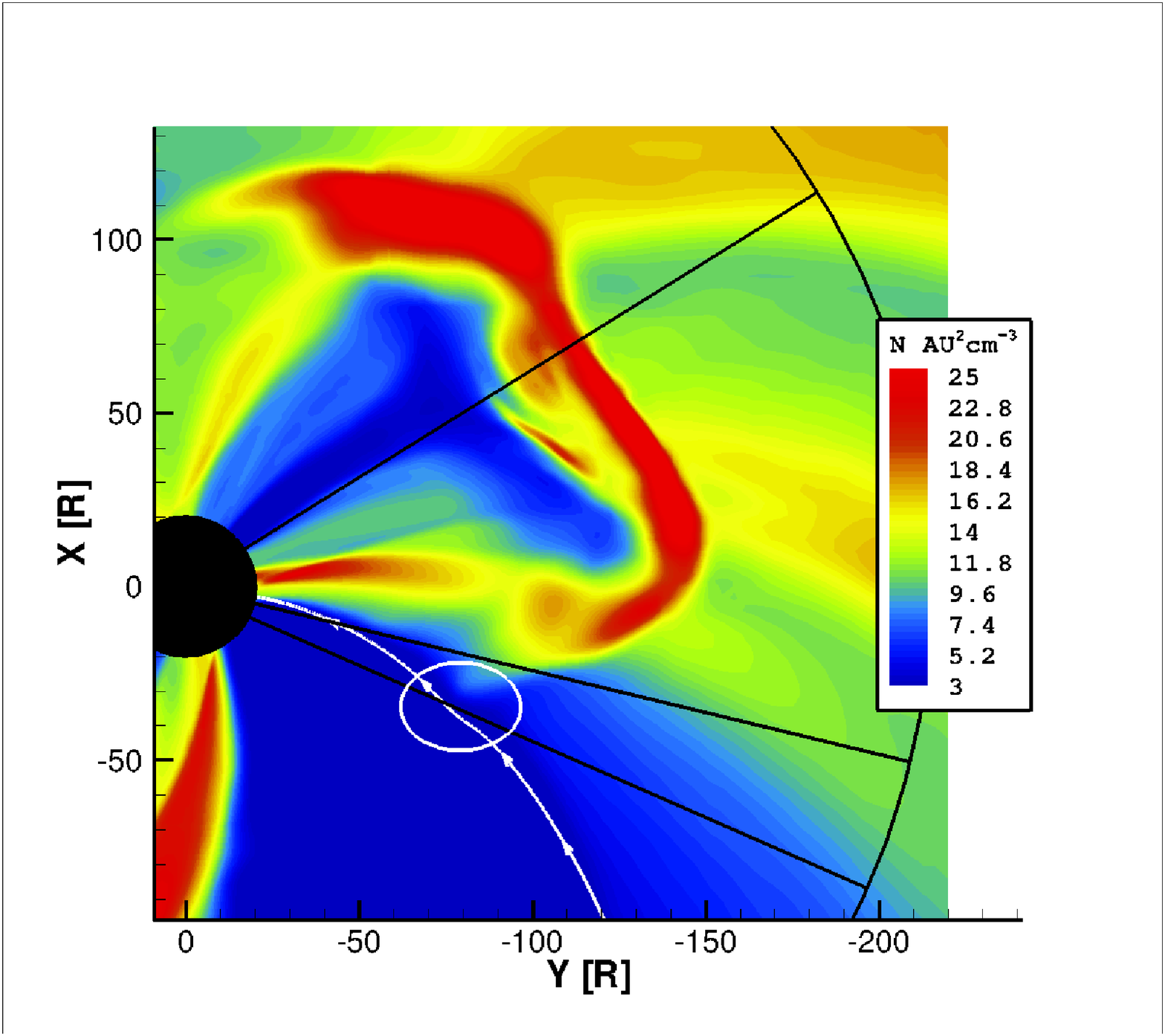}} 
{\includegraphics*[width=6.6cm]{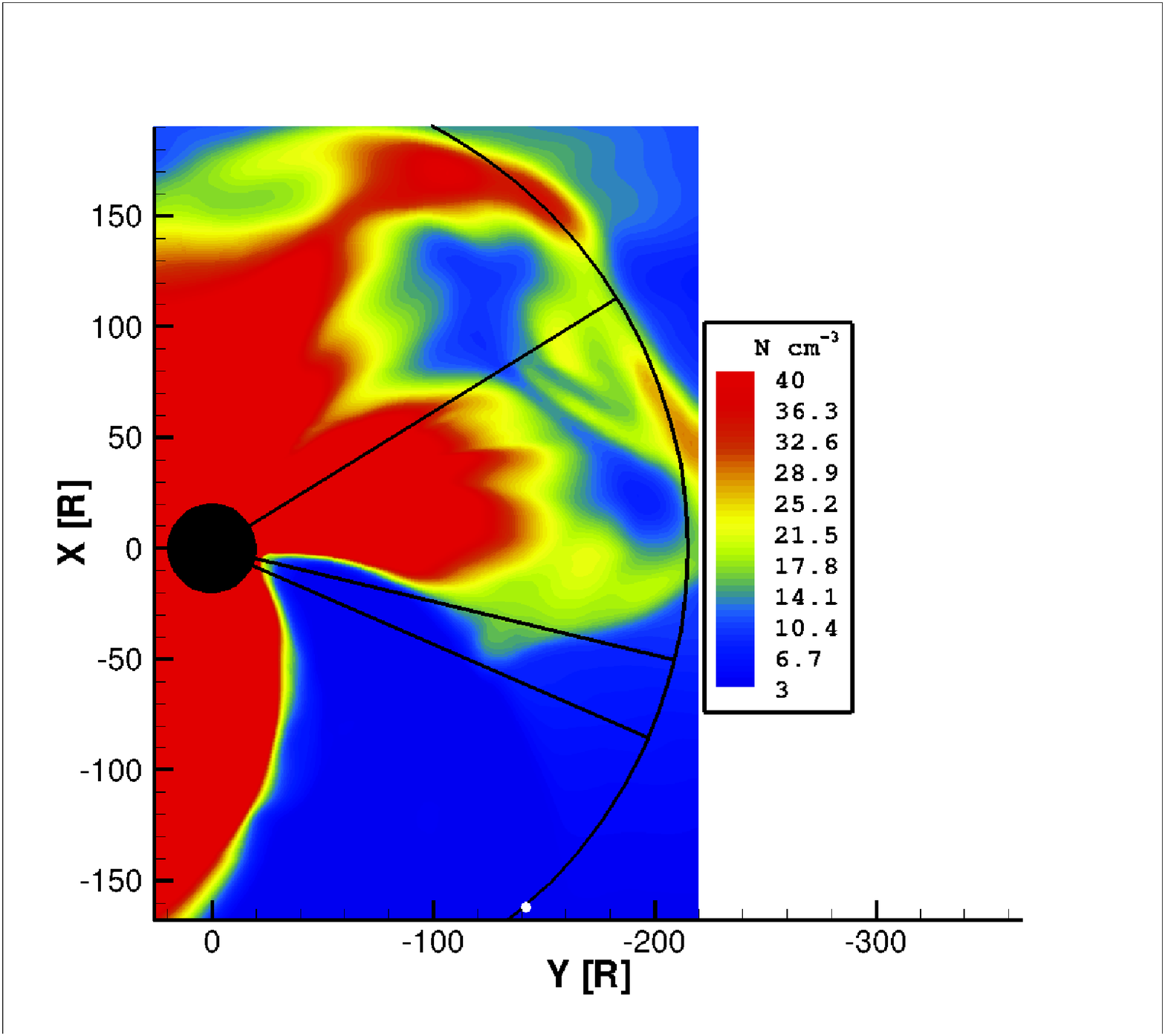}}
\end{center}
\end{minipage}\hfill
\caption{Number density ({\it left} scaled by 1/R$^2$) in the equatorial plane 29 hours ({\it left}) and 48 hours ({\it right}) after the start of the shearing phase. The black circle represent Earth's orbit, Earth's position is shown with a white disk. The three lines are, from top to bottom the radial directions corresponding to the central position of the eruption, the longitude reached at 58 hours and the maximum longitudinal extent. The field line connecting to Earth is shown in white and the shock position is highlighted with the white ellipse.}
\label{fig:OneAU}
\end{figure}

As described above, we find that, by the end of the shearing phase, the erupting flux has reconnected with open magnetic field lines from AR~67 through the null point and some erupting field line are now direclty connected to Earth's vicinity. This can be seen in the top right panel of Figure~2. The disruption of these field lines due to the passage of the shock wave can be seen at the later times. We find that the shock wave has been formed by 5~$R_{\odot}$ (within the first hour after the start of the shearing phase). It has also been significantly deflected due to reconnection at the northern null point. An approximate visualization of the shock wave can be seen on the left panel of Figure~3 along with the field lines connecting to Earth's vicinity. We should also note that the shock does not become quasi-parallel along Earth-connected field lines until about 90~minutes after the start of the eruption \citep[see also][]{Roussev:2008}. A similar evolution of the shock angle (from quasi-perpenicular to quasi-parallel) with distance has been previously reported in \citet{Manchester:2005} for a field line about 37$^\circ$ north of the center of the CME.

\section{Heliospheric Evolution and Results at 1~AU}

A shock wave associated with the eruption of August 24 was detected at Earth by {\it ACE} spacecraft about 58~hours after the start of the eruption. In our simulation, the shock wave does not extend all the way to Earth. In fact, we find that the shock wave reaches 1~AU approximatively 42 hours after the end of the shearing phase about 60$^\circ$ West of Earth. Its maximum angular span is about 120$^\circ$ and it reaches a maximum of 25$^\circ$ West of Earth. After 58~hours, the simulated shock wave reaches a point about 35$^\circ$ West of Earth. The results at 48 hours are shown on the right panel of Figure~4. 

In agreement with previous studies \citep[e.g.][]{Jacobs:2007}, we find that there is no significative non-radial expansion of the CME in the heliosphere past the upper corona. Although in our simulation the shock wave does not hit Earth, we find that there is direct magnetic connection between the flank of the shock and Earth during most of the heliospheric evolution of the CME and its associated shock. This is illustrated at time 29 hours in the left panel of Figure~4 and may have important consequences to understand and predict the observed Forbush decrease of ACRs \citep[]{Eroshenko:2008} and the time variation of the SEP event.

\section{Discussions and Conclusion}
We performed a Sun-to-Earth simulation of the August 24, 2002 CME event with a realistic CME initiation mechanism \citep[]{Roussev:2007}. This is one of the first solar-terrestrial simulation involving a CME model more complex than a flux rope or a simple perturbation added onto the solar wind. Using a realistic model is required to study space weather effect such as (i) the formation of the shock wave, (ii) the change of connectivity between the Earth and the solar surface during the eruption, and (iii) the possible deflection of the CME. We find that reconnection of the erupting flux at one null point north-east of the active region results in the deflection of the eruption in the corona (we find no subsequent deflection in the heliosphere) towards the Sun-Earth line. There is an opening of part of the erupting flux due to this reconnection event, and, consequently, a change of magnetic connectivity with Earth. We find that the shock wave has formed by 5~$R_\odot$ and has a sufficient longitudinal extent to accelerate particles along Earth-connected field lines. 

Last, we must investigate why, in our simulation, the shock does not reach Earth. First, contrary to what was inferred by \citet{Wang:2004}, we find no consequent deflection of the CME in the heliosphere, even though it is a fast western CME. This might be because the solar wind is not well reproduced. Or it might be because there is in fact no significant deflection of CME in the heliosphere. If we believe that there is no large deflection in the heliosphere, then the shock angular extent must be at least 170$^\circ$ to explain the detection of the shock wave by {\it ACE}, significantly larger than what is predicted by our model. It is worth noting that the simulated CME is slower in the corona than the observed one by as much as 35$\%$; a faster CME will most likely be associated with a larger shock wave. We will investigate this in future simulations of the same event. In situ observations by STEREO (and Helios) and future polar coronagraphs will provide valuable information concerning the angular extent of shock wave, and possible deflection. 

\acknowledgments
  The research for this manuscript was supported by NSF grants ATM0639335 
  and ATM0819653 and NASA grants NNX07AC13G and NNX08AQ16G.

\end{document}